\documentclass{llncs}
\usepackage{llncsdoc}
\usepackage{times}
\usepackage{url}
\usepackage{helvet}
\usepackage{courier}
\usepackage{amsmath}
\usepackage{amsfonts}
\usepackage{amssymb}
\usepackage{color}
\usepackage{bbm}
\usepackage{graphicx}
\usepackage{verbatim}
\usepackage{tabularx}
\usepackage{graphicx}      
\usepackage{times}

\usepackage{url}
\usepackage{algorithm, algorithmicx, algpseudocode}
\usepackage{subfig}

\newcommand\blfootnote[1]{%
  \begingroup
  \renewcommand\thefootnote{}\footnote{#1}%
  \addtocounter{footnote}{-1}%
  \endgroup
}

\begin{document}

%
\title{On the Erd\H{o}s Discrepancy Problem*}

\author{Ronan Le Bras \and Carla P. Gomes \and Bart Selman}

\institute{Computer Science Department\\ Cornell University, Ithaca NY 14850}

\maketitle

\blfootnote{*Submitted on April 14, 2014 to the 20th International Conference on Principles and Practice of Constraint Programming.}

\begin{abstract}
According to the Erd\H{o}s discrepancy conjecture, for any infinite $\pm 1$ sequence, there exists a homogeneous arithmetic progression of unbounded discrepancy. In other words, for any $\pm 1$ sequence $(x_1,x_2,...)$ and a discrepancy $C$, there exist integers $m$ and $d$ such that $|\sum_{i=1}^m x_{i \cdot d}| > C$. This is an $80$-year-old open problem and recent development proved that this conjecture is true for discrepancies up to $2$. Paul Erd\H{o}s also conjectured that this property of unbounded discrepancy even holds for the restricted case of completely multiplicative sequences (CMSs), namely sequences $(x_1,x_2,...)$ where $x_{a \cdot b} = x_{a} \cdot x_{b}$ for any $a,b \geq 1$. The longest CMS with discrepancy $2$ has been proven to be of size $246$. In this paper, we prove that any completely multiplicative sequence of size $127,646$ or more has discrepancy at least $4$, proving the Erd\H{o}s discrepancy conjecture for CMSs of discrepancies up to $3$. In addition, we prove that this bound is tight and increases the size of the longest known sequence of discrepancy $3$ from $17,000$ to $127,645$. Finally, we provide inductive construction rules as well as streamlining methods to improve the lower bounds for sequences of higher discrepancies.

\end{abstract}

\section{Introduction}
\label{introduction}
Discrepancy theory addresses the problem of distributing points uniformly over some geometric object, and studies how irregularities inevitably occur in these distributions. For example, this subfield of combinatorics aims to answer the following question: for a given set $U$ of $n$ elements, and a finite family $\mathcal{S}=\{S_1,S_2,\dots,S_m\}$ of subsets of $U$, is it possible to color the elements of $U$ in red or blue, such that the difference between the number of blue elements and red elements in any subset $S_i$ is small? 

Important contributions in discrepancy theory include the Beck-Fiala theorem \cite{beck1981integer} and Spencer's Theorem \cite{spencer1985six}. The Beck-Fiala theorem guarantees that if each element appears at most $t$ times in the sets of $\mathcal{S}$, the elements can be colored so that the imbalance, or discrepancy, is no more than $2t-1$. According to the Spencer's theorem, the discrepancy of $\mathcal{S}$ grows at most as $\Omega( \sqrt{n \log (2m/n)})$. Nevertheless, some important questions remain open.

According to Paul Erd\H{o}s himself, two of his oldest conjectures relate to the discrepancy of homogeneous arithmetic progressions (HAPs) \cite{erdos1982some}. Namely, a HAP of length $k$ and of common difference $d$ corresponds to the sequence $(d, 2d, \dots, kd)$. The first conjecture can be formulated as follows:

\begin{conjecture}\label{conj1}
Let $(x_1,x_2,...)$ be an arbitrary $\pm 1$ sequence. The discrepancy of $x$ w.r.t. HAPs must be unbounded, i.e. for any integer $C$ there is an integer $m$ and an integer $d$ such that $|\sum_{i=1}^m x_{i \cdot d}| > C$.
\end{conjecture}

This problem has been open for over eighty years, as is the weaker form according to which one can restrict oneself to completely multiplicative functions.  Namely, $f$ is a completely multiplicative function if $f(a \cdot b)=f(a) \cdot f(b)$ for any $a, b$. The second conjecture translates to:


\begin{conjecture}\label{conj2}
Let $(x_1,x_2,...)$ be an arbitrary completely multiplicative $\pm 1$ sequence. The discrepancy of $x$ w.r.t. HAPs must be unbounded, i.e. for any integer $C$ there is a $m$ and a $d$ such that $|\sum_{i=1}^m x_{i \cdot d}| > C$.
\end{conjecture}

Hereinafter, when non-ambiguous, we refer to the discrepancy of a sequence as its discrepancy with respect to homogeneous arithmetic progressions. Formally, we denote $disc(x) = max_{m,d} |\sum_{i=1}^m x_{i \cdot d}|$.
We denote $\mathcal{E}_1(C)$ the length for which any sequence has discrepancy at least $C+1$, or equivalently, one plus the maximum length of a sequence of discrepancy $C$. Similarly, we define $\mathcal{E}_2(C)$ the length for which any \emph{completely multiplicative} sequence has discrepancy at least $C+1$. \footnote{Note that, if Conjecture \ref{conj1} (resp. Conjecture \ref{conj2}) were to be rejected, $\mathcal{E}_1(C)$ (resp.  $\mathcal{E}_2(C)$ ) would correspond to infinity.}

A proof or disproof of these conjectures would constitute a major advancement in combinatorial number theory \cite{nikolov2013hereditary}. To date, both conjectures have been proven to hold for the case $C \leq 2$. The values of $\mathcal{E}_1(1),\mathcal{E}_2(1)$, and $\mathcal{E}_2(2)$ have been long proven to be $12, 10$, and $247$ respectively, while recent development proved $\mathcal{E}_1(2)=1161$ \cite{konev2014sat}. Konev and Lisitsa \cite{konev2014sat} also provide a new lower bound for $\mathcal{E}_1(3)$. After 3 days of computation, a SAT solver was able to find a satisfying assignment for a sequence of length $13,000$. Yet, it would fail to find a solution of size $14,000$ in over 2 weeks of computation. They also report a solution of length $17,000$, the longest known sequence of discrepancy $3$. 
In this paper, we substantially increase the size of the longest sequence of discrepancy $3$, from $17,000$ to $127,645$. In addition, we claim that $\mathcal{E}_2(3)=127,646$, making this bound tight, as \texttt{Plingeling} was able to prove unsat and \texttt{Lingeling} generated an UNSAT proof in DRUP format \cite{heule2013trimming}. 

This paper is organized as follows. The next section formally defines the Erdos discrepancy problems (for the general case and the multiplicative case) and presents SAT encodings for both problems. We then investigate streamlined search techniques to boost the search for lower bounds of these two problems, and to characterize additional structures that appear in a subset of the solutions. Furthermore, in a subsequent section, we provide construction rules that are based on these streamliners and allow to generate larger sequences of limited discrepancy from smaller ones. The last section presents the results of these approaches. 

%

\section{Problem Formulation}
\label{encoding}
In this section, we first formally define the two conjectures as decision problems and then propose encodings for these problems.

\begin{definition}[$\textbf{EDP}_1$]
Given two integers $n$ and $C$, does there exist a $\pm 1$ sequence $(x_1,\dots,x_n)$ such that $| \sum_{i=1}^m x_{i \cdot d}| \leq C$ for any $1 \leq d \leq n, m \leq n/d$. 
\end{definition}

Konev and Lisitsa \cite{konev2014sat} provide a SAT encoding for this problem that uses an automaton accepting any sub-sequence of discrepancy exceeding $C$. A state $s_j$ of the automaton corresponds to the sum of the input sequence, while the accepting state $s_B$ captures whether the sequence has exceeded the discrepancy $C$. A proposition $s_j^{(m,d)}$ is true whenever the automaton is in state $\sum_{i=1}^{m-1} x_{i \cdot d}$ after reading the sequence $(x_d,\dots,x_{(m-1)d})$. Let $p_i$ be the proposition corresponding to $x_i = +1$. A proposition that tracks the state of the automaton for an input sequence $(x_d,x_{2d},\dots,x_{\lfloor n/d \rfloor d})$ can be formulated as:

\begin{align}
\phi(n,C,d)=s_0^{(1,d)} \bigwedge_{m=1}^{n/d} \bigg(    &\bigwedge_{-C \leq j < C} \big(s_j^{(m,d)} \land p_{id} \to s_{j+1}^{(m+1,d)} \big)  \wedge \notag\\  &\bigwedge_{-C < j \leq C} \big(s_j^{(m,d)} \land \overline{p_{id}} \to s_{j+1}^{(m+1,d)} \big) \wedge \notag\\ & \big( s_C^{(m,d)} \land p_{id} \to s_B  \big)    \wedge                                        \notag\\ &\big( s_{-C}^{(m,d)} \land \overline{p_{id}} \to s_B  \big) \bigg)
\end{align}

In addition, we need to encode that the automaton is in exactly one state at any point in time. Formally, we define this proposition as:

\begin{equation}
\chi(n,C) = \bigwedge_{1\leq d \leq n/C, 1\leq m \leq n/d} \bigg(   \bigvee_{-C\leq j \leq C} s_j^{(i,d)}    \wedge  \bigwedge_{-C\leq j_1,j_2 \leq C} \big( \overline{s}_{j_1}^{(i,d)} \vee \overline{s}_{j_2}^{(i,d)}   \big)             \bigg)
\end{equation}

Finally, we can encode the Erd\H{o}s Discrepancy Problem as follows:
\begin{equation}
\textbf{EDP}_1(n,C):  \overline{s}_B \wedge \chi(n,C) \wedge \bigwedge_{d=1}^{n} \phi(n,C,d)
\end{equation}

Furthermore, as the authors of \cite{konev2014sat}, the actual states $s_j^{(m,d)}$ of the automaton do not require $2C+1$ binary variables to represent the $2C+1$ values of the states. Instead, one can modify this formulation and use $\left\lceil log_2(2C+1) \right\rceil$ binary variables to encode the automaton states.

For the completely multiplicative case, we introduce additional constraints to capture the multiplicative property of any element of the sequence, i.e. $x_{id}=x_i x_d$ for any $1 \leq d \leq n, 1 \leq i \leq n/d$. With respect to the boolean variables $p_i$, $p_d$ and $p_{id}$, such a constraint acts as XNOR gate of input $p_i$ and $p_d$ and of output $p_{id}$. Formally, we denote this proposition $\mathcal{M}(i,d)$ and define:

\begin{equation}
\mathcal{M}(i,d)= (p_i \vee p_d \vee p_{id}) \wedge  (\overline{p_i} \vee \overline{p_d} \vee p_{id})  \wedge  (p_i \vee \overline{p_d} \vee \overline{p_{id}})  \wedge  (\overline{p_i} \vee p_d \vee \overline{p_{id}})  
\end{equation}

Importantly, for completely multiplicative sequences, the discrepancy of the subsequence ($x_d,...,x_{md}$) of length $m$ and common difference $d$  will be the same as the discrepancy of the subsequence ($x_1,...,x_{m}$). Indeed , we have $|\sum_{i=1}^m x_{i \cdot d}|=|\sum_{i=1}^m x_{i} x_{d}|= |x_{d}|\cdot|\sum_{i=1}^m x_{i}|=|\sum_{i=1}^m x_{i}|$. Therefore, one needs only check that the partial sums $\sum_{i=1}^m x_{i}, 1\leq m\leq n$ never exceed $C$ nor go below $-C$.
Furthermore, note that a completely multiplicative sequence is entirely characterized by the values it takes at prime positions, i.e. $\{x_p | p \text{ is prime}\}$. In addition, if there exists a completely multiplicative sequence sequence $(x_1,...,x_{p-1})$ of discrepancy $C$ with $p$ prime, then the sequence  $(x_1,...,x_{p-1},$ $(-1)^{\mathbbm{1}_{\sum_{i=1}^m x_{i} \geq 0}})$ will also be a CMS of discrepancy $C$. As a result, $\mathcal{E}^2(C)$ cannot be a prime number.

Overall, for the completely multiplicative case, we obtain:
\begin{equation}
\textbf{EDP}^2(n,C):  \overline{s}_B \wedge \chi(n,C) \wedge \phi(n,C,1) \bigwedge_{1 \leq d \leq n, 1 \leq i \leq n/d} \mathcal{M}(i,d)
\end{equation}

\section{Streamlined Search}
\label{streamlining}
The encoding of \textbf{EDP}$_1$ given in the previous section has successfully led to prove a tight bound for the case $C=2$ \cite{konev2014sat}. On an Intel Core i5-2500K CPU, it takes about 800 seconds for \texttt{Plingeling} \cite{biere2013lingeling} to find a satisfying assignment for \textbf{EDP}$_1(1160,2)$ and less than 6 hours for \texttt{Glucose} \cite{audemard2013glucose} to generate a proof of $\mathcal{E}_1(2)=1,161$. Nevertheless, for the case $C=3$, it requires more than 3 days of computation for \texttt{Plingeling} to find a sequence of size $n=13,000$, and fails to find a sequence of size $14,000$ in over two weeks of computation. 

In this section, in order to improve this lower bound and acquire a better understanding of the solution space, we explore streamlining techniques that identifies additional structure occurring in a subset of the solutions. Among the solutions of a combinatorial problem, there might be solutions that possess regularities beyond the structure of the combinatorial problem itself. 
Streamlining \cite{gomes04} is an effective combinatorial search strategy that exploits these additional regularities. By intentionally imposing additional structure to a combinatorial problem, it focuses the search on a highly structured subspace and triggers increased constraint reasoning and propagation. This search technique is sometimes referred to as ``tunneling'' \cite{kouril2005resolutiontunnels}. In other words, a streamlined search consists in adding specific desired or observed regularities, such as a partial pattern that appears in a solution, to the combinatorial solver. These additional regularities boost the solver that may find more effectively larger solutions that contain these regularities. If no solution is found, the observed regularities were likely accidental. Otherwise, one can analyze these new solutions and suggests new regularities. This methodology has been successfully applied to find efficient constructions for different combinatorial objects, such as spatially-balanced Latin squares \cite{LeBras2011FromStreamlined}, or graceful double-wheel graphs \cite{LeBras2013Double}.

When analyzing solutions of \textbf{EDP}$_1(n,2)$ for $n \in [1,1160]$, there is a feature that visually stands out of the solutions. When looking at a solution as a $2D$-matrix with entries in $\{-,+\}$ and changing the dimensions of the matrix, there seems to be clear preferred matrix dimensions (say $m$-by-$p$) such that the $m$ rows are mostly identical for the columns $1$ to $p-1$, suggesting that $x_i = x_{i \bmod{p}}$ for $1 \leq i \leq p-1$. We denote $period(x,p,t)$ the streamliner that enforces this observation and define:
\begin{equation}
period(x,p,t): x_i = x_{i\bmod p} ~\forall 1 \leq i \leq t, i\not \equiv 0\bmod{p}
\end{equation}

First, while this observation by itself did not allow to improve the current best lower bound for $\mathcal{E}_1(3)$, it led to the formulation of the construction of the next section. Second, it also led to the re-discovery of the so-called 'improved Walters sequence' \cite{WaltersSeq}, defined as follows:
\begin{equation}
\mu_3(i) = \begin{cases} +1, & \mbox{if } i\mbox{ is }1 \bmod 3 \\ -1, & \mbox{if } i\mbox{ is } 2 \bmod 3 \\ -\mu_3(i / 3), & \mbox{otherwise.} \end{cases}
\end{equation}

In the following, we denote $walters(x,w)$ the streamliner imposing that the first $w$ elements of a sequence $x$ follow the improved Walters sequence, i.e.:
\begin{equation}
walters(x,w): x_i = \mu_3(i) ~\forall 1 \leq i \leq w
\end{equation}

One can easily see that the improved Walters sequence is a special case of the periodic sequence defined previously. Namely, for any sequence $x$ where $walters(x,w)$ holds true, then we have $period(x,9,w)$.

Finally, another striking feature of the solutions of \textbf{EDP}$_1(n,2)$ is that they tend to follow a multiplicative sequence. Interesting, \textbf{EDP}$_2$ restricts \textbf{EDP}$_1$ to the special case of multiplicative functions and we observe for the case $C=2$ that this restriction substantially impacts the value of the best bound possible (i.e. $\mathcal{E}_1(2)=1,161$ whereas $\mathcal{E}_2(2)=247$). Nevertheless, the solutions of \textbf{EDP}$_1(n,2)$ exhibit a partial multiplicative property and we define:
 \begin{equation}
mult(x,m,l): x_{i\cdot d} = x_i x_d ~\forall 2 \leq d \leq m, 1 \leq i \leq n/d, i \leq l
\end{equation}

In the experimental section, we show the speed-ups that are triggered using these streamliners, and how the best lower bound for \textbf{EDP}$_1(n,2)$ gets greatly improved.

\section{Construction Rule}
\label{construction}
In this section, we show how we used insights from the $period(x,p,t)$ streamliner in order to generate an inductive construction rule for sequences of discrepancy $C$ from sequences of lower discrepancy. 

Consider a sequence $x$ that is periodic of period $p$, as defined in the previous section, i.e. $period(x,p,|x|)$ holds true, and is of length $n=p*k$. Then, the sequence $x$ can be written as: 
\begin{align}
x = ( &y_1, y_2, \ldots, y_{p-2}, y_{p-1}, z_1\notag\\
&y_1, y_2, \ldots, y_{p-2}, y_{p-1}, z_2\notag\\
&\ldots\notag\\
&y_1, y_2, \ldots, y_{p-2}, y_{p-1}, z_k) \label{eqx}
\end{align}

Let $C$ be the discrepancy of $z=(z_1, z_2,...,z_k)$ and $C'$ the discrepancy of $(y_1,...,y_{p-1})$. Given that $\sum_{i=1}^m x_{ip} = \sum_{i=1}^m z_{i}$ for any $1 \leq m \leq k$, we have $disc(x) \geq C$. Note that if $x$ was completely multiplicative, then it would hold $disc(x) = C$. We study the general case where $x$ is not necessarily multiplicative, and investigate the conditions under which $disc(x)$ is guaranteed to be less or equal to $C+C'$.

For a given common difference $d$ and length $m$, we consider the subsequence $(x_d,x_{2d},...,x_{md})$. Let $q=\frac{p}{gcd(d,p)}$. Given the definition \ref{eqx} of $x$, the subsequence $(x_d,x_{2d},...,x_{md})$ corresponds to:
\begin{align}
(&y_{d \bmod{p}},y_{2d \bmod{p}},...,y_{(q-1)d \bmod{p}},z_{q},\\&y_{d \bmod{p}},y_{2d \bmod{p}},...,y_{(q-1)d \bmod{p}},z_{2q},\\&y_{d \bmod{p}},...)
\end{align}

Note that if $p$ divides $d$ or $d$ divides $p$, this subsequence becomes $(z_q,z_{2q},...,z_{qm})$ and is of discrepancy at most $C$. 
As a result, a sufficient condition for $x$ to be of discrepancy at most $C+C'$ is to have $y_{d \bmod{p}},y_{2d \bmod{p}},...,y_{(q-1)d \bmod{p}}$ of discrepancy $C'$ and summing to $0$. We say that such a sequence has a discrepancy $\bmod{p}$ of $C'$. Formally, we define the problem of finding such sequences as follows:

\begin{definition}[Discrepancy $\bmod{p}$]
Given two integers $p$ and $C'$, does there exist a $\pm 1$ sequence $(y_1,\dots,y_{p-1})$ such that:
\begin{align}
| \sum_{i=1}^m y_{i \cdot d \bmod{p}}| \leq C',&~~\forall 1 \leq d \leq n, m < \frac{p}{gcd(d,p)}\\
\sum_{i=1}^{\frac{p}{gcd(d,p)}-1} y_{i \cdot d \bmod{p}} = 0,&~~\forall 1 \leq d \leq n \label{eqsum0}
\end{align}
\end{definition}

Notice that, given the equation \ref{eqsum0}, $p$ should be odd for such a sequence to exist.

We encode this problem as a Constraint Satisfaction Problem (CSP) in a natural way from the problem definition. We provide the experimental results in the next section.

\section{Results}
\label{results}
All experiments were run on a Linux (version 2.6.18) cluster where each node has an Intel Xeon Processor X5670, with dual-CPU, hex-core @2.93GHz, 12M Cache, 48GB RAM. Unless otherwise noted, the results were obtained using the parallel SAT solver \texttt{Plingeling}, version \texttt{ats1} for the SAT encodings, and using \texttt{IBM ILOG CPLEX CP Optimizer}, release \texttt{12.5.1} for the CP encodings.

First, we evaluate the proposed streamliners for the two problems. Table \ref{tab:res} reports the length of the sequences that were successfully generated, as well as the computation time. The first clear observation is that, for EDP$_1$, the streamlined search based on the partial multiplicative property significantly boosts the search and allows to generate solutions that appear to be out of reach of the standard search approach. For example, while it takes about 10 days to find a solution of length $13,900$ without streamliners, the streamlined search generates a substantially-large satisfying assignment of size $31,500$ in about 15 hours.
Next, we study streamliners that were used for EDP$_2$, i.e. partially imposing the walters sequence. The results clearly show the speed up triggered by the combination of the new encoding for EDP$_2$ with the \emph{walters} streamliners. Interestingly, the longest \emph{walters} sequence of discrepancy $3$ is of size $819$. Nevertheless, one can successfully impose the first $800$ elements of the walters sequence and still expand it to a sequence of length $108,000$.
Furthermore, when imposing $walters(730)$, it takes less than 1 hour and an half to find a satisfying assignment for a sequence of size $127,645$. Moreover, without additional streamliners, it takes about 60 hours to prove unsat for the case $127,646$ and allows us to claim that this bound is tight. Nevertheless, the solver generates a DRUP proof of size 335GB, which lies beyond the reach of traditional checkers \cite{heule2013trimming}.

\begin{table*}[ht]
\centering
\small
\begin{tabular}{cccc}
			\hline
			\\[-1.5ex]
			\emph{Encoding} & \emph{Streamliners} &  \emph{Size of sequence} & \emph{Runtime (in sec)}
			\\[1pt]
			\hline \\[-2ex]
			 & - & 13,000 & 286,247 \\
			 & - & 13,500 & 560,663\\
 			& - & 13,900 & 770,122\\
			& mult(120,2000) & 15,600 & 4,535\\
			EDP$_1$ & mult(150,2000) & 18,800 & 8,744\\
			 & mult(200,1000)  & 23,900 & 12,608\\
			& mult(700,10000) & 27,000 & 45,773\\
			& mult(700,20000) & 31,500 & 51,144\\[1pt]
			\hline \\[-2ex]
			& walters(800) & 81,000 & 1,364\\
			EDP$_2$ & walters(800) & 108,000 & 4,333\\
			 & walters(700)  & 112,000 & 5,459\\
			& walters(730) & 127,645 & 4,501\\[1pt]
			\hline \\[-2ex]

		\end{tabular}
\caption{Solution runtimes of searches with and without streamliners. The streamlined search leads to new lower bounds for the 2 EDP problems.}
\label{tab:res}
\end{table*}

In terms of the inductive construction described in the previous section, we can generate sequences whose discrepancy $\bmod{p}$  is $1$, for $p$ in $1, 3, 5, 7,$ and $9$, while it also generates sequences of discrepancy $\bmod{p}$ equal to $2$ for $p$ in $11, 13, 15, 17, 25, 27, 45$, and $81$. Overall, this proves that one can take any sequence $x$ of length $|x|$ and discrepancy $C$ and generate one of length $9|x|$ and of discrepancy $C+1$, or of length $81|x|$ and of discrepancy $C+2$. As a result, this provides a new bound for the case of discrepancy $4$, and proves $\mathcal{E}_1(4) > 9*127645= 1,148,805$. Interestingly, such a long sequence suggests that the proof of the Erdos conjecture for $C>3$ may require additional insights and analytical proof, beyond the approach proposed in this work.


\section{Conclusions}
\label{conclusions}
In this paper, we address the Erdos discrepancy problem for general sequences as well as for completely multiplicative sequences. We adapt a SAT encoding previously proposed and investigate streamlining methods to speed up the solving time and understand additional structures that occur in some solutions. Overall, we substantially improve the best known lower bound for discrepancy 3 from $17,001$ to $127,646$. In addition, we claim that this bound is tight, as suggested by the unsat proof generated by \texttt{Lingeling}. Finally, we propose construction rules to inductively generate longer sequences of limited discrepancy.

\section{Acknowledgments}
\label{acknowledgments}
This work was supported by the National Science Foundation (NSF IIS award, grant 1344201). The experiments were run on an infrastructure supported by the NSF Computing research infrastructure for Computational Sustainability grant (grant 1059284).


\bibliographystyle{splncs}
\bibliography{erdosdiscrepancy}


\end{document}